\documentclass[9pt,twocolumn,twoside]{osajnl}
\usepackage{scalerel}
\usepackage{tikz}
\usetikzlibrary{svg.path,calc}

\definecolor{orcidlogocol}{HTML}{A6CE39}
\tikzset{
   orcidlogo/.pic={
    \fill[orcidlogocol] svg{M256,128c0,70.7-57.3,128-128,128C57.3,256,0,198.7,0,128C0,57.3,57.3,0,128,0C198.7,0,256,57.3,256,128z};
    \fill[white] svg{M86.3,186.2H70.9V79.1h15.4v48.4V186.2z}
                 svg{M108.9,79.1h41.6c39.6,0,57,28.3,57,53.6c0,27.5-21.5,53.6-56.8,53.6h-41.8V79.1z M124.3,172.4h24.5c34.9,0,42.9-26.5,42.9-39.7c0-21.5-13.7-39.7-43.7-39.7h-23.7V172.4z}
                 svg{M88.7,56.8c0,5.5-4.5,10.1-10.1,10.1c-5.6,0-10.1-4.6-10.1-10.1c0-5.6,4.5-10.1,10.1-10.1C84.2,46.7,88.7,51.3,88.7,56.8z};
  }
}

\newcommand\orcidicon[1]{\href{https://orcid.org/#1}{\mbox{
\begin{tikzpicture}[overlay,remember picture]
\coordinate (A);
\coordinate(B) at ($(A)-(2pt,-9pt)$);
\end{tikzpicture}
\begin{tikzpicture}[overlay,remember picture,yscale=-0.045,xscale=0.045,transform shape]
\pic at (B) {orcidlogo};
\end{tikzpicture}
}{}}}

\usepackage{hyperref}
 
\journal{ol}
\setboolean{shortarticle}{true}

\title{Noise in supercontinuum generated using PM and non-PM tellurite-glass all-normal dispersion fibers}

\author[1,*\protect\orcidicon{0000-0003-2007-0930}\,\,]{Shreesha Rao D. S.}
\author[2\protect\orcidicon{0000-0002-1707-2129}\,\,]{Tanvi Karpate}
\author[3,4\protect\orcidicon{0000-0002-5297-8321}\,\,]{Amar Nath Ghosh}
\author[1\protect\orcidicon{0000-0002-2618-0631}\,\,]{Iv\'{a}n B. Gonzalo}
\author[2\protect\orcidicon{0000-0002-3110-9792}\,\,]{Mariusz~Klimczak}
\author[5]{Dariusz Pysz}
\author[2,5\protect\orcidicon{0000-0003-2863-725X}\,\,]{Ryszard Buczy\'{n}ski}
\author[3]{Cyril Billet}
\author[1,6\protect\orcidicon{0000-0002-8041-9156}\,\,]{Ole Bang}
\author[3\protect\orcidicon{0000-0001-9520-9699}\,\,]{John~M.~Dudley}
\author[3\protect\orcidicon{0000-0001-5849-1533}\,\,]{Thibaut Sylvestre}

\affil[1]{DTU Fotonik, Department of Photonics Engineering, Technical University of Denmark, \O rsteds Plads, 2800 Kongens Lyngby, Denmark}
\affil[2]{University of Warsaw, Faculty of Physics, Pasteura 7, 02-093 Warsaw, Poland}
\affil[3]{Institut FEMTO-ST, CNRS, UMR 6174, Universit\'{e} Bourgogne Franche-Comt\'{e}, Besan\c{c}on, France}
\affil[4]{Optoelectronics Research Centre, University of Southampton, Southampton, SO17 1BJ, United Kingdom}
\affil[5]{\L{}ukasiewicz Research Network - Institute of Microelectronics and Photonics, Al. Lotnik\'{o}w 32/46, Warsaw, 02-668, Poland}
\affil[6]{NKT Photonics A/S, Blokken 84, 3460 Birker\o d, Denmark}

\affil[*]{Corresponding author: shreesharaods@gmail.com}

\begin{abstract}
Intensity fluctuations in supercontinuum generation are studied in polarisation-maintaining (PM) and non-PM all-normal dispersion tellurite photonic crystal fibers. Dispersive Fourier transformation was used to resolve shot-to-shot spectra generated using 225~fs pump pulses at 1.55~$\mu$m, with experimental results well reproduced by vector and scalar numerical simulations. By comparing the relative intensity noise for the PM and non-PM cases, supported by simulations, we demonstrate the advantage of the polarisation-maintaining property of the PM fibers in preserving low-noise dynamics. We associate the low-noise in the PM case with the suppression of polarisation modulation instability.
\end{abstract}

\setboolean{displaycopyright}{true}

\begin{document}

\maketitle

Motivated by important applications that require low-noise broadband light sources, there has been a recent interest in supercontinuum (SC) generation in all-normal dispersion (ANDi) fiber~\cite{Anu21Per,Shr21Shot,Kob21Near,Thi21Rec}. In addition to providing a flat-top and broadband spectrum, SC generated in ANDi fibers can yields high shot-to-shot coherence due to the relative insensitivity to input noise of self-phase modulation and optical wave breaking dynamics in the normal dispersion regime. The superior noise properties of ANDi SC over SC generated by pumping in the anomalous dispersion have been verified by measurements of spectral fluctuations using unequal path Michelson interferometers, RF beating with stabilised laser diodes, relative intensity noise (RIN), and dispersive Fourier transformation (DFT)~\cite{Klm16DFT,Nis18Cha,Rao19ULow,Tar19CompDft}. However, it has also been established that SC coherence in ANDi fibers can degrade with increased pulse duration and fiber length due to parametric interaction between coherent and incoherent components~\cite{Ale17Lim}, as well as polarisation modulation instability (PMI)~\cite{IvnRIN18}.

Experiments have used both polarisation-maintaining (PM) and weakly birefringent (non-PM) ANDi fibers to demonstrate low-noise SC generation. With PM-ANDi fiber, low-noise SC can be obtained by matching the linear pump polarisation to one of the principal axes of the fiber~\cite{Tar19CompDft,Alex20Pmi,Eti20Cro}. With non-PM-ANDi fiber, the onset of PMI depends on the pump pulse duration, fiber length, and peak power (P$_0$)~\cite{IvnRIN18}. As a result, very specific pump parameters and fiber lengths need to be chosen to avoid PMI-induced noise, limiting the spectral broadening that can be achieved~\cite{Rao19ULow}. 

A difficulty with these previous studies, however, is that the range of experimental parameters used makes it difficult to quantitatively compare the SC properties obtained using PM and non-PM ANDi fiber. In this Letter, we address this problem directly through a combined experimental and numerical study of SC noise in PM and non-PM ANDi fibers under controlled conditions. Specifically, we use two variants of a highly nonlinear ANDi tellurite glass photonic crystal fiber (PCF); one fabricated with polarisation-maintaining functionality, the other without. Both variants share a hexagonal air-hole lattice structure. The DFT technique is used to quantify the RIN of SC generated in both fibers, and we explicitly show the superior noise properties of the SC when the PM-ANDi fiber variant is used. These experimental results are supported by numerical simulations using both scalar and coupled generalized nonlinear Schr\"{o}dinger equation (GNLSE) models.

\begin{figure}[htbp!]
\centering
\includegraphics[width=\linewidth]{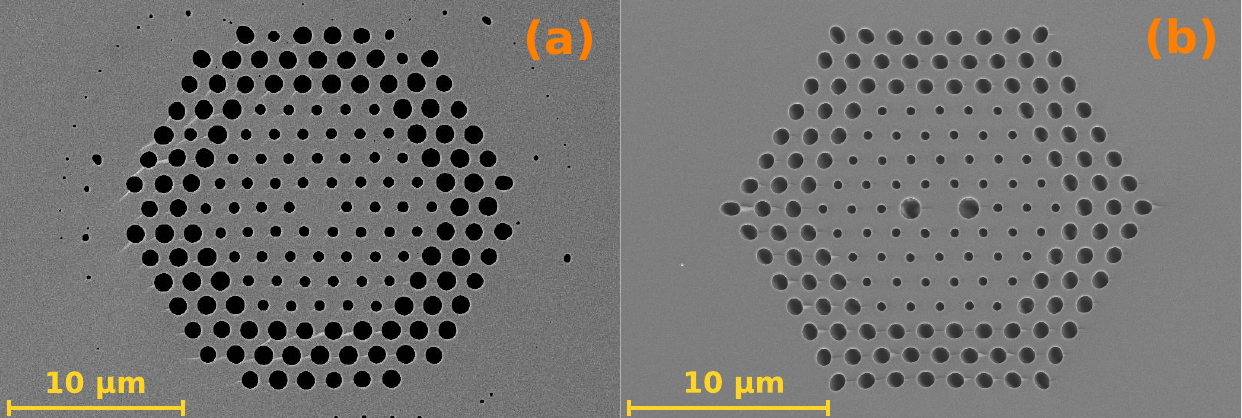}
\includegraphics[width=\linewidth]{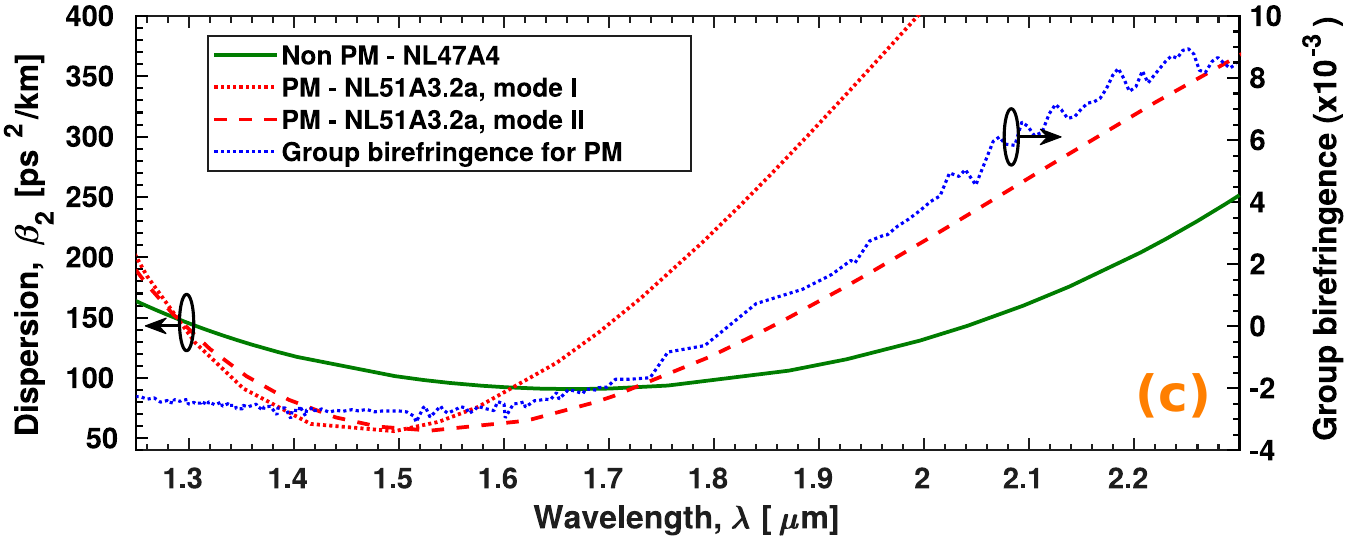}
\caption{SEM images of \textbf{(a)} non-PM and \textbf{(b)} PM fibers. \textbf{(c)} Left axis: GVD of the non-PM fiber (solid line) and the two fundamental modes of the PM fiber – mode I (dotted line) and mode II (dashed line). Right axis: PM fiber’s group birefringence (dash–dotted line).}
\label{fig:DispPmANpm}
\rule{\linewidth}{0.81pt}
\end{figure}
The non-PM fiber (labelled NL47A4) was previously studied for its SC performance in Ref.~\cite{Kli19TNpm}. This fiber was used as the starting point to develop the PM variant, (labelled NL51A3.2). Scanning electron microscope (SEM) images of both fibers are shown in Fig.~\ref{fig:DispPmANpm}(a-b). The measured group velocity dispersion (GVD) of the non-PM, and the two degenerate modes of the PM fibers are plotted on the left axis of Fig.~\ref{fig:DispPmANpm}(c). The right axis of Fig.~\ref{fig:DispPmANpm}(c) plots the measured group birefringence of the PM fiber.

Pulses with a full width half maximum, T$_{FWHM}=$ 225~fs from an optical parametric oscillator (Coherent Chameleon compact) at 1.55 $\mu$m and a repetition rate of 80.15~MHz were used to generate SC in both fibers. A 40$\times$ microscope objective was used for coupling, and a half-wave plate (Thorlabs AQWP10M-1600) was used to control the polarisation orientation. 

The SC output was coupled into a 170~m long dispersion-shifted fiber (DSF) for shot-to-shot spectral measurements using DFT \cite{Wet12Dft1,God13Dft2,Goda13Dft}. The input to the DSF was attenuated to ensure linear propagation. The DSF had a normal GVD of $\beta_2=$107 ps$^2$/km, and a dispersion slope $\beta_3=$0.082 ps$^3$/km at 1.55 $\mu$m. The DFT setup used a 50~GHz InGaAs detector (u$^2$t: XPDV2120R) and a 12~GHz, 40~GS/s oscilloscope (Agilent DSA91204A). The pulse train was recorded for 5 $\mu$s, corresponding to 400 consecutive SC pulses. The 5.8~nm spectral resolution of the DFT was limited by the system bandwidth. 

\begin{figure}[htbp!]
\centering
\includegraphics[width=\linewidth]{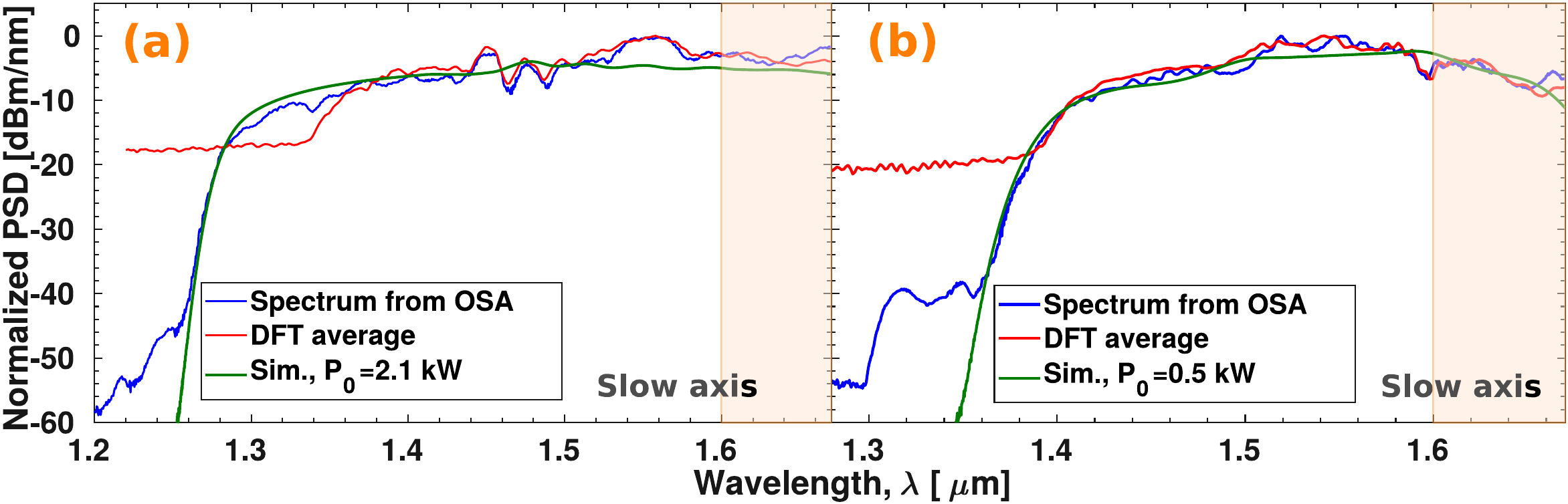}
\caption{Spectra measured using the OSA (blue), the DFT average (red), and from simulations when light is coupled along the slow axes (green): \textbf{(a)} for the non-PM fiber using vector simulations with P$_0=$2.1 kW; and \textbf{(b)} for the PM fiber using scalar simulations with P$_0=$0.5 kW.}
\label{fig:DftAnSpec}
\rule{\linewidth}{1pt}
\end{figure}
Figures~\ref{fig:DftAnSpec}(a-b) show the average SC spectra measured using DFT (red) for P$_0=$ 2.1 kW for a 12.9 cm long non-PM fiber and P$_0=$ 0.5 kW for a 12.6 cm long PM fiber. Power spectral density (PSD) measured using an optical spectrum analyzer (OSA) (Agilent 86142B) are plotted in blue, and the simulated spectra in green (simulation details are given later). The DFT and OSA spectra agree well for wavelengths longer than 1.35 $\mu$m but there is poorer agreement at shorter wavelengths because of the effect of DSF dispersive terms higher than $\beta_3$. However, this short-wavelength mismatch does not affect the conclusions that can be drawn. We also note that the reduced sensitivity of the InGaAs detector and lower SC power above 1.6~ $\mu$m results in unreliable RIN measurements. Hence, we limit our discussion to wavelengths below 1.6~$\mu$m, with longer wavelengths (the shaded region) shown only for reference.

\begin{figure}[htpb!]
\centering
\includegraphics[width=\linewidth]{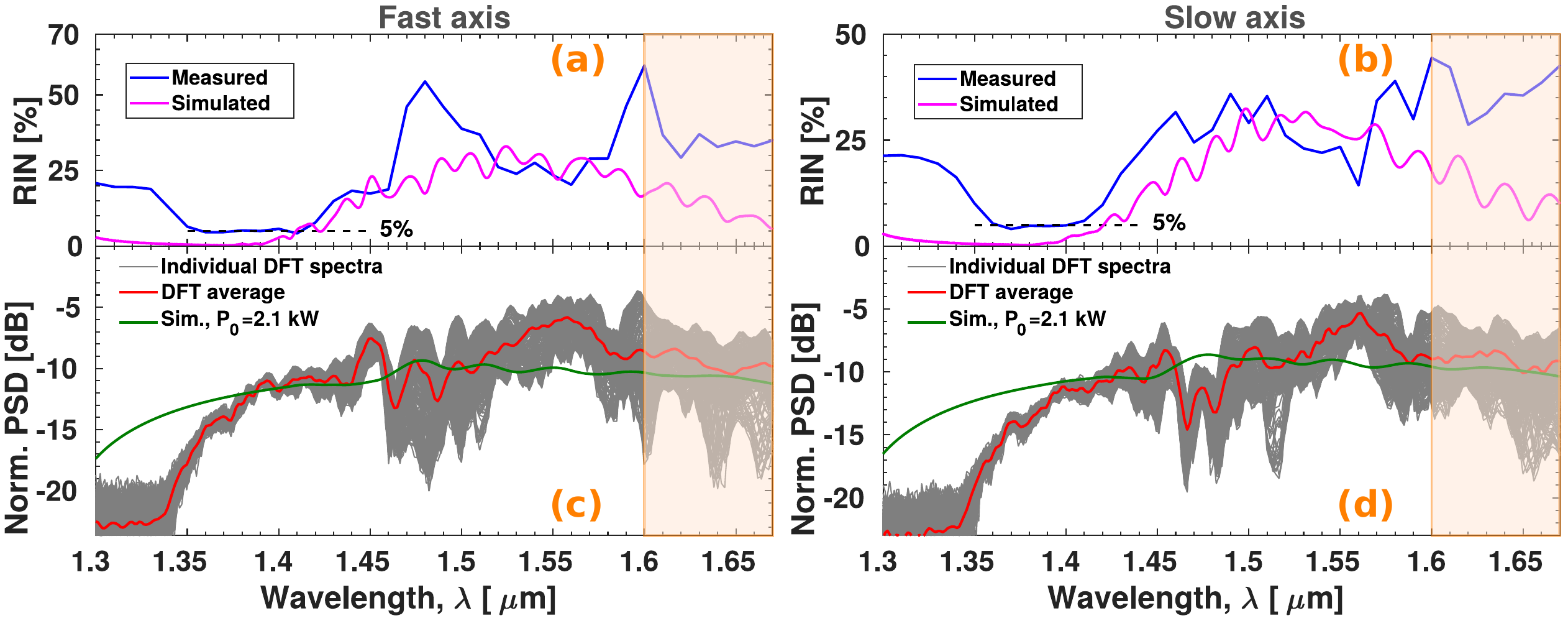}
\includegraphics[width=\linewidth]{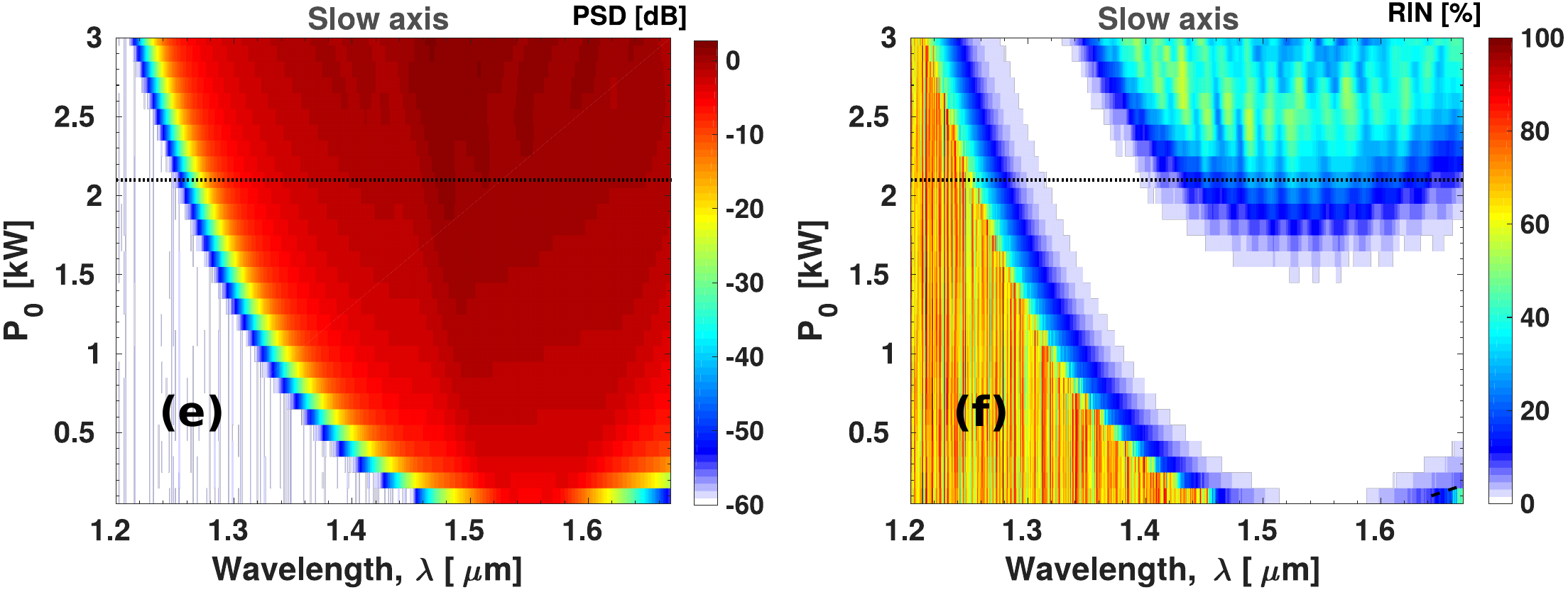}
\caption{\textbf{Results for the non-PM ANDi fiber}: \textbf{(a)} measured and simulated SC RIN when light is coupled into the fast axis, and \textbf{(b)} when light is coupled into the slow axis. The measured 400 individual DFT spectra (grey) and the DFT average (red) and simulated SC for P$_0=$ 2.1 kW (green), when light is coupled into the fast axis \textbf{(c)}, and when light is coupled into the slow axis \textbf{(d)}. SC spectra from CGNLSE simulations with P$_0$ varied from 0.1 kW to 3 kW (forty power increments) when light is coupled into the slow axis \textbf{(e)} and the corresponding RIN plots \textbf{(f)}.}
\label{fig:DftNonPmFP}
\rule{\linewidth}{1pt}
\end{figure}

Power was coupled into the non-PM ANDi fiber's fast axis. The DFT average (red) is plotted in Fig.~\ref{fig:DftNonPmFP}(c). The grey lines are the 400 individual DFT spectra. RIN was calculated as the ratio of standard deviation to the mean of the 400 DFT spectra and is shown in Fig.~\ref{fig:DftNonPmFP}(a). The input polarisation into the non-PM ANDi fiber was then rotated by 90$^\circ$ to couple into the slow axis of the fiber. Figure~\ref{fig:DftNonPmFP}(d) shows the DFT average of the spectrum (red) and the 400 individual pulses (grey), and the RIN of the SC is shown in Fig.~\ref{fig:DftNonPmFP}(b). The two fundamental axes of the fiber were identified by observing the polarisation of the light leaving the fiber at a low power while the input polarisation to the fiber was varied. At the powers used in the experiments, the axis at which the broadest SC was generated is denoted as the fast axis, as the dispersion for the pump pulse when coupled to this axis is lower than the dispersion when coupled to the slow axis. 

CGNLSE~\cite{Agr12NoFi} was used to numerically simulate the SC and the noise properties of the SC in the non-PM ANDi fiber. The effective refractive index of the fiber was numerically calculated using the fiber parameters, and the full dispersion profile of the fiber was used in the simulations. The mode profile dispersion is included in the simulation such that the photon number is conserved when the net loss is zero~\cite{Lag07Mde}. The total loss of the fiber was included in the simulation and was calculated as the numerically obtained confinement loss and the fiber material loss measured in Ref.~\cite{Kli19TNpm}. The weak birefringence between the two fundamental modes was accounted for by their phase mismatch, $\Delta \beta= \Delta n~\omega_0 /c$, where $\omega_0$ is the angular frequency of the pump. The implementation of the CGNLSE is similar to that in Ref.~\cite{IvnRIN18}. Nonlinear index, n$_2=$ 4.88$\times$10$^{-19}$ m$^2$W$^{-1}$, birefringence, $\Delta n=$ 10$^{-7}$, and the Raman response curve with single Lorentzian profile with damping time of vibrations, $\tau_1=$5.5 fs, $\tau_1=$32 fs, and fractional contribution of the delayed Raman response, $f_R=$0.2 were used in the simulations. The input pulse was Gaussian with T$_{FWHM}=$225 fs. Quantum noise in the pump was included by adding independent and normally distributed real and imaginary parts in each time bin with the width $\Delta t$ of the input envelope function in the Wigner representation. The quantum noise has a variance of $\hbar\omega_0/2\Delta t$~\cite{Corney01QN2,ZhBa2016Noi}. The pump was measured to have a RIN $=$ 0.6\%, using the DFT technique. This technical laser noise was added to the P$_0$ of the input pulse such that the energy P$_0$T$_{FWHM}$ was constant~\cite{Rao19ULow,Gen19Amp}. All RIN computations were done with an ensemble of twenty independent simulations with different noise seeds, and all simulated spectra presented in the figures were averaged over those twenty individual simulations. 

\begin{figure}[htbp!]
\centering
\includegraphics[width=\linewidth]{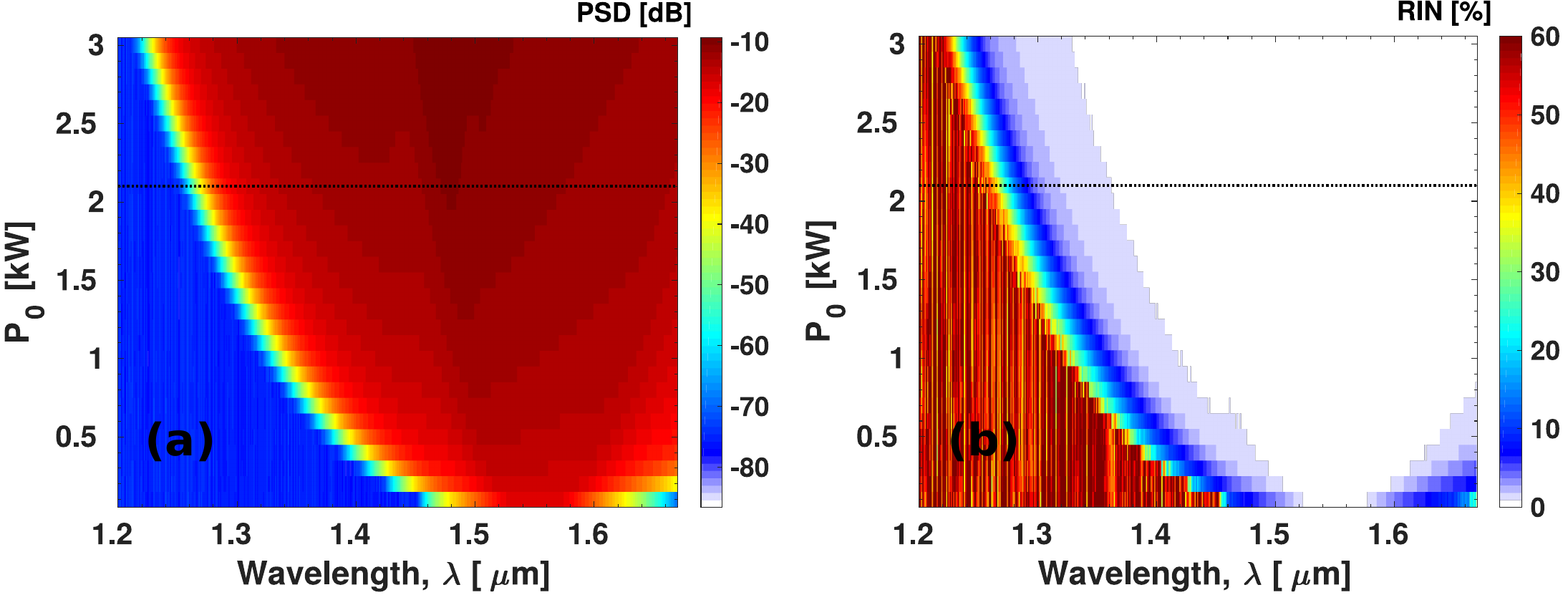}
\caption{\textbf{Numerical results for the non-PM ANDi fiber}: \textbf{(a)} Colormap of SC obtained using scalar-GNLSE simulations with increasing P$_0$ from 0.1 kW to 3 kW (a set of forty simulations); and \textbf{(b)} the corresponding numerical RIN traces.}
\label{fig:SGnWRnPM}
\rule{\linewidth}{1pt}
\end{figure}

The spectrum obtained from the CGNLSE simulations (green) when coupled into the slow axis of the non-PM fiber with P$_0=$2.1 kW is plotted in Fig.~\ref{fig:DftAnSpec}(a), along with the experimentally measured spectrum (blue) and the measured DFT average (red). The CGNLSE simulation reproduces the measured spectrum very well. The spectrum obtained from the CGNLSE simulations (green) for the slow axis with P$_0=$2.1 kW is plotted in Fig.~\ref{fig:DftNonPmFP}(d), as well, along with the 400 individual shots in grey and the measured DFT average, for refernce. 
Fig.~\ref{fig:DftNonPmFP}(b) shows the corresponding RIN plot. The blue curve is the RIN calculated from the experimental DFT data, and the magenta curve is the RIN calculated from the CGNLSE simulations. The RIN obtained from the experimental DFT data exceeds 5\% [indicated by dashed lines in Figs.~\ref{fig:DftNonPmFP}(a-b)] over the entire SC width and exceeds 50\% at wavelengths longer than 1.4 $\mu$m [when coupled to the slow axis, as shown in Fig.~\ref{fig:DftNonPmFP}(b)]. The amplitudes of the DFT-measured shot-to-shot fluctuations match the RIN profile obtained from the simulations. Specifically, the simulation follows well the slope at wavelengths where the RIN begins to rise, at the shorter wavelength side. Similarly, Fig.~\ref{fig:DftNonPmFP}(c) shows the measured and simulated spectra for the fast axis, while Fig.~\ref{fig:DftNonPmFP}(a) shows the corresponding RIN curves. Figure~\ref{fig:DftNonPmFP}(e) shows SC from forty sets of CGNLSE simulations with P$_0$ varied from 0.1 kW to 3 kW for the slow axis. The dotted horizontal line marks the spectrum generated with P$_0=$2.1 kW, corresponding to the P$_0$ used in the experiment. The corresponding RIN plotted in Fig.~\ref{fig:DftNonPmFP}(f) demonstrates that an input P$_0$ greater than 1.5 kW causes large pulse-to-pulse fluctuations in the output SC. Figures~\ref{fig:DftNonPmFP}(a-f) show that when a weakly birefringent ANDi fiber is used along with a relatively long pump pulse and a large absolute value of dispersion, the SC generation dynamics exhibit large pulse-to-pulse fluctuations, in most of the cases due to PMI, as shown in Ref.~\cite{IvnRIN18}. To verify the origin of the large pulse-to-pulse fluctuations observed in the non-PM fiber, we implemented a separate scalar-GNLSE simulation in the interaction picture~\cite{Hult07For} with all the input parameters the same as in the CGNLSE simulations. The implementation of the scalar- GNLSE is similar to that in Ref.~\cite{Rao19ULow}, and further details on the implementation can be found there. We carried out forty sets of scalar-GNLSE simulations with P$_0$ varied from 0.1 kW to 3 kW for the non-PM fiber. The spectrum at the end of the fiber is plotted in Fig.~\ref{fig:SGnWRnPM}(a) and the corresponding RIN plots are shown in Fig.~\ref{fig:SGnWRnPM}(b). 
The dotted horizontal lines in Figs.~\ref{fig:SGnWRnPM}(a-b) mark the P$_0=$2.1 kW used in the experiments. The RIN plots in Fig.~\ref{fig:SGnWRnPM}(b) show that the pulse-to-pulse fluctuations in the SC caused by mixed parametric Raman (MPR) noise alone are low for all of the P$_0$ in the plot, indicating that the high RIN observed in Fig.~\ref{fig:DftNonPmFP}(f) is indeed caused by PMI. 

\begin{figure}[htbp!]
\centering
\includegraphics[width=\linewidth]{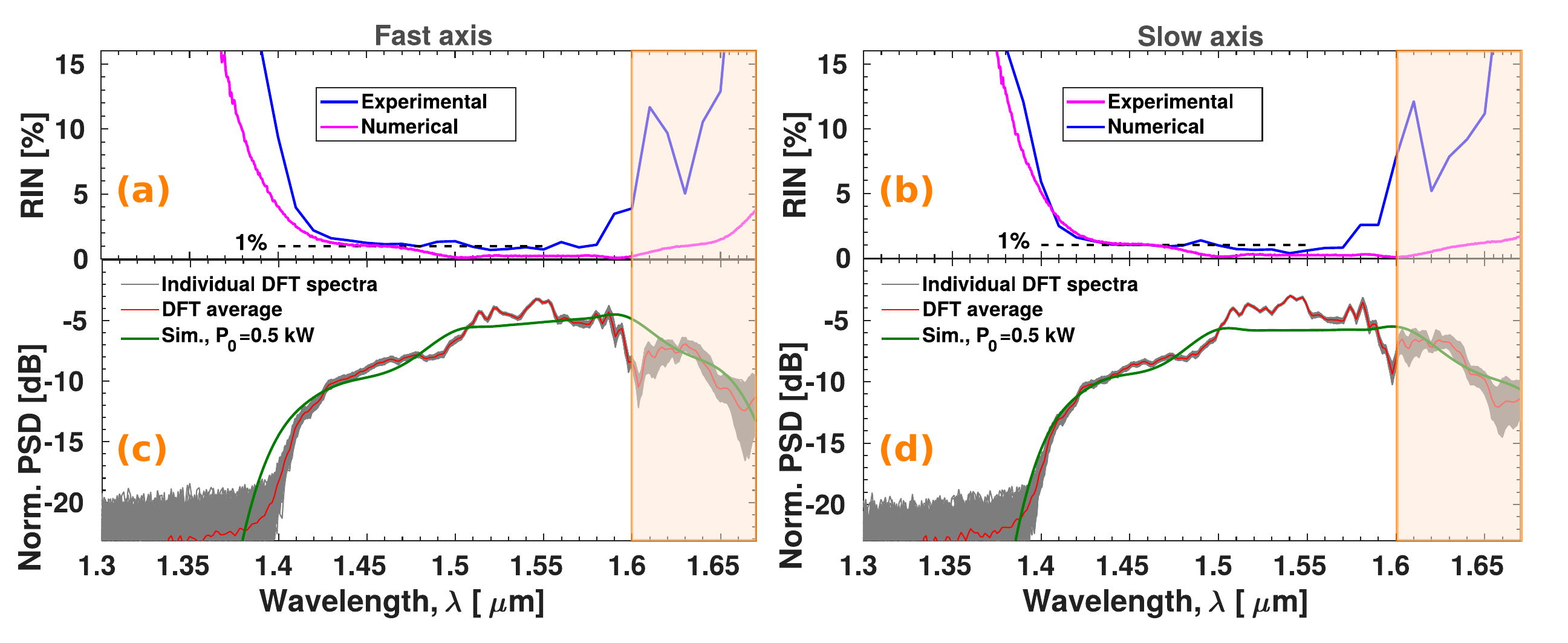}
\includegraphics[width=\linewidth]{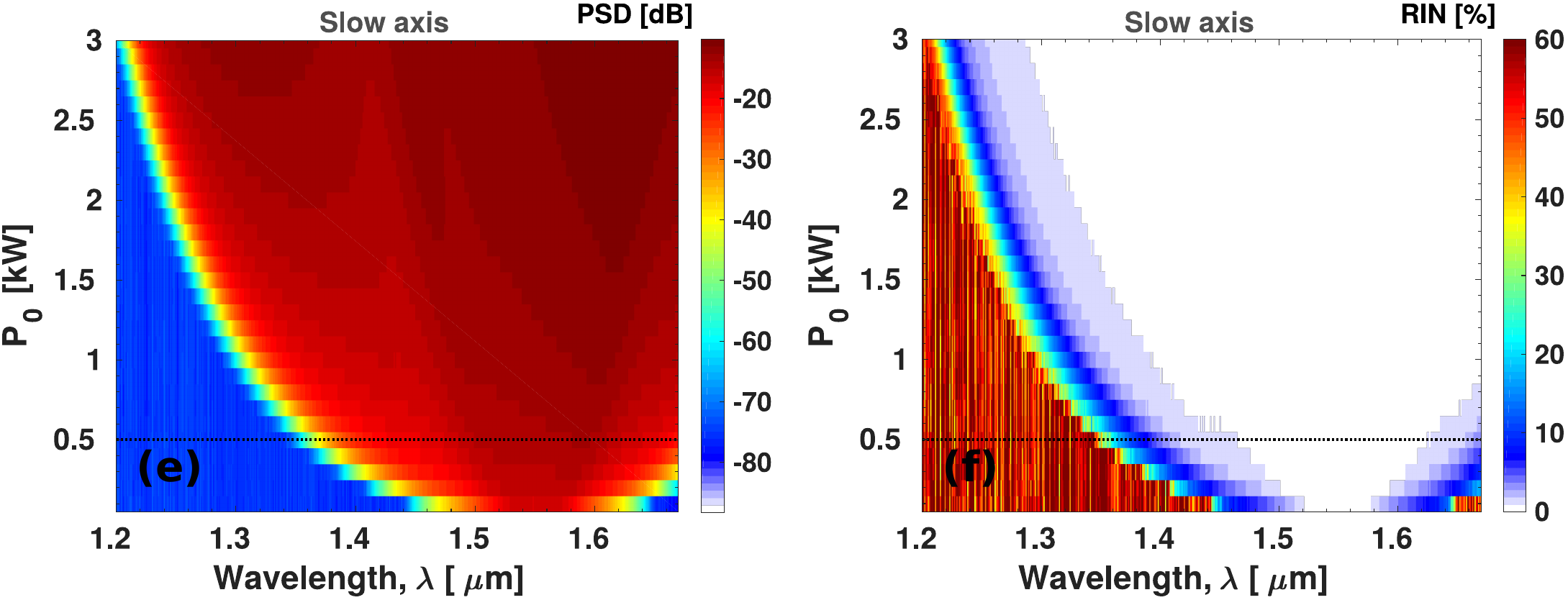}
\caption{\textbf{Results for the PM ANDi fiber}: \textbf{(a)} measured and simulated RIN of the SC out of the fiber when light is coupled into the fast axis, and \textbf{(b)} when light is coupled into the slow axis. The measured 400 individual DFT spectra (grey) and the DFT average (red) and simulated SC for P$_0=$ 0.5 kW (green), when light is coupled into the fast axis \textbf{(c)}, and when light is coupled into the slow axis \textbf{(d)}. SC at the end of the fiber obtained using scalar GNLSE simulations with P$_0$ varied from 0.1 kW to 3 kW (a set of forty simulations) when light is coupled into the slow axis \textbf{(e)}, and the corresponding RIN plots \textbf{(f)}.}
\label{fig:DfPmFP}
\rule{\linewidth}{1pt}
\end{figure}

Next, we investigated the noise characteristics of the SC using the PM variant of the fiber. The pump polarisation was first aligned to the fast axis of the PM ANDi fiber. The measured DFT average (red) is plotted in Fig.~\ref{fig:DfPmFP}(c). The grey traces are the overlaying spectra of the 400 SC shots resolved using the DFT. The polarisation of the pump was then rotated by 90$^\circ$ to couple into slow axis of the fiber. Figure~\ref{fig:DfPmFP}(d) shows the measured DFT average (red) and the 400 individual spectra (grey) for this case. The SC and the RIN out of the fiber were simulated using scalar-GNLSE. The spectrum obtained from the scalar-GNLSE simulations (green) with P$_0=$0.5 kW is plotted in Fig.~\ref{fig:DfPmFP}(d). The spectrum obtained from the scalar-GNLSE simulations (green) for the slow axis with P$_0=$0.5 kW is plotted in Fig.~\ref{fig:DftAnSpec}(b), along with the experimentally measured spectrum (blue) and the measured DFT average (red), for reference. We see that the scalar-GNLSE simulation agrees very well with the measured spectrum. Figure~\ref{fig:DfPmFP}(b) shows the corresponding RIN plots. The blue curve is the experimental RIN calculated using the DFT spectra. The measured RIN is low and is around 1\% [indicated by dashed lines in Figs.~\ref{fig:DfPmFP}(a-b)] across the entire SC width where our DFT measurements are accurate. The magenta curve is the RIN numerically calculated from the simulations. The measured shot-to-shot fluctuations, characterised in terms of RIN, agree very well with the RIN obtained from the simulations. Specifically, the simulation follows well the low-RIN measured in the experiments for wavelengths shorter than 1.6 $\mu$m, where our measurements are accurate. Similarly, Fig.~\ref{fig:DfPmFP}(c) shows the measured and simulated spectra for the fast axis, while Fig.~\ref{fig:DfPmFP}(a) shows the corresponding RIN curves. 
Fig.~\ref{fig:DfPmFP}(e) shows SC from forty sets of scalar-GNLSE simulations with P$_0$ varied from 0.1 kW to 3 kW for the slow axis. The dotted horizontal line marks the spectrum generated with P$_0=$0.5 kW, corresponding to the P$_0$ used in the experiments. The corresponding RIN plots in Fig.~\ref{fig:DfPmFP}(f) show that for all the P$_0$ values used in the simulation, the RIN of the SC leaving the PM ANDi fibers remains extremely low.

The CGNLSE simulations for the non-PM fiber and the scalar-GNLSE simulations for the PM fiber are able to accurately reproduce the results obtained in the experiments. This allows us to confidently use the simulation to compare the noise properties of the two fibers. It is noted that even though the same pump configuration (with the pump set to its maximum output power) was used in the experiments with the PM fiber, the power coupled into the PM fiber was lower than that into the non-PM fiber. This is attributed to the presence of the large holes introduced to induce birefringence, which leads to a higher mode mismatch between the focused beam and the fundamental modes of this fiber. Despite the difference in P$_0$ coupled into the PM and non-PM fibers, the noise properties of the two can be directly compared because all other experimental parameters were the same.

A DFT noise study in Ref.~\cite{Klm16DFT} used a weakly birefringent all-solid soft glass ANDi fiber, pumped by a 390 fs laser pulse at 1.55 $\mu$m. Unlike in our case, the noise observed was attributed to MPR noise and PMI was not associated with the observed pulse-to-pulse fluctuations. The SC out of a germanium doped PM ANDi fiber, pumped by a 25 fs laser pulse at 1.55 $\mu$m was studied in Ref.~\cite{Tar19CompDft}. Even though the individual DFT spectra seemed relatively stable, the pulse-to-pulse noise quantified as SNR (which is the inverse of RIN) shows large variations across the SC spectrum. These results individually do not provide a clear picture of the conditions needed to obtain a low-noise SC. Therefore, we have studied the noise properties of the SC for both the PM and the non-PM ANDi fibers, and we have experimentally verified them in the relatively long fs pulse regime. The RIN plot for the SC from the non-PM fiber [see Fig.~\ref{fig:DftNonPmFP}(f)] shows that for a low pump P$_0$, the RIN remains low, but then rises due to PMI as P$_0$ increases. However, for the same pulse duration and the set of P$_0$ value, the SC out of the PM fiber always remains low-noise [see Fig.~\ref{fig:DfPmFP}(f)], when pumped along its fundamental axis.

In conclusion, we have experimentally characterised spectral fluctuations in SC generation from a weakly birefringent non-PM ANDi fiber and its PM version when pumped by 225~fs pulses at 1.55 $\mu$m. The experimentally measured RIN was reproduced very well using CGNLSE simulations for the non-PM fiber and using scalar-GNLSE simulations for the PM fiber. We show that for the non-PM fiber, PMI plays a detrimental role while attempting to maintain low-noise, and for the fiber and pump parameters used in our case, MPR noise has a negligible effect. Finally, using the same pump conditions, we have demonstrated that it is indeed possible to obtain ultra-low-noise (RIN$<$1\%) SC when pumping along the fundamental axis of the PM version of the ANDi fiber.

\begin{backmatter}
\bmsection{Funding} Horizon 2020 Framework Programme Marie Curie grant No. (722380 [SUPUVIR]); Villum Fonden (2021 Villum Investigator: Table-Top Synchrotrons); Fundacja na rzecz Nauki Polskiej (FNP) (First TEAM/2016-1/1 - POIR.04.04.00-00-1D64/16-00); Agence Nationale de la Recherche (ANR) (ANR-15-IDEX-0003,
ANR-17-EURE-0002,ANR-21-ESRE-0040).

\bmsection{Disclosures} The authors declare no conflicts of interest.

\bmsection{Data availability} Data underlying the results presented in this paper are not publicly available at this time but may be obtained from the authors upon reasonable request.
\end{backmatter}
\bibliography{DftNoiAndiRef}
\end{document}